\documentclass[prb,superscriptaddress]{revtex4-1}

\usepackage{epsfig}
\usepackage{amsmath}

\newcommand{\oos}{$\omega_i \omega_j s_{ij}$}
\newcommand{\odo}{$\nu\omega_i \nabla^2 \omega_i$}
\newcommand\ome{$\mbox{\boldmath{$\omega$}}$}

\newcommand\lai{$\mbox{\boldmath{$\lambda_i$}}$}
\newcommand\laa{$\mbox{\boldmath{$\lambda_1$}}$}
\newcommand\lab{$\mbox{\boldmath{$\lambda_2$}}$}
\newcommand\lac{$\mbox{\boldmath{$\lambda_3$}}$}

\newcommand\vis{$\nabla^2 \mbox{\boldmath{$\omega$}}$}

\begin{document}

\setlength{\baselineskip}{1.6\baselineskip}

\noindent\textbf{\large Viscous tilting and production of
vorticity in homogeneous turbulence}

\medskip
\hspace{1cm}

\noindent M. Holzner$^1$, M. Guala$^2$, B. L\"{u}thi$^1$, A.
Liberzon$^3$, N. Nikitin$^4$, W. Kinzelbach$^1$ and A.
Tsinober$^{3}$

\noindent\textit{\small $^1$ Institute of Environmental Engineering, ETH Zurich, CH 8093 Zurich, Switzerland}%
\newline
\textit{\small $^2$ Galcit, California Institute of Technology,
Pasadena CA 91125, USA}
\newline
\textit{\small $^3$ School of Mechanical Engineering, Tel Aviv
University, Ramat Aviv 69978, Israel}%
\newline
\textit{\small $^4$ Institute of Mechanics, Moscow State
University, 119899 Moscow, Russia}\\%
\vspace{1em}\hspace{8em} (International Collaboration for
Turbulence Research)


\date{\today}

\begin{abstract}
Viscous depletion of vorticity is an essential and well known
property of turbulent flows, balancing, in the mean, the net
vorticity production associated with the vortex stretching
mechanism. In this letter we however demonstrate that viscous
effects are not restricted to a mere destruction process, but play
a more complex role in vorticity dynamics that is as important as
vortex stretching. Based on results from particle tracking
experiments (3D-PTV) and direct numerical simulation (DNS) of
homogeneous and quasi isotropic turbulence, we show that the
viscous term in the vorticity equation can also locally induce
production of vorticity and changes of its orientation (viscous
tilting).
\end{abstract}

\pacs{}

\maketitle


In turbulent flows, the energy is injected at large scales by some
forcing mechanism and dissipated into heat through the effect of
viscosity at the smallest scales of motion, e.g.
Ref.~[\onlinecite{tennekes}]. The main physical mechanisms that
control fluid turbulence at the smallest scales are commonly
described in terms of strain and vorticity, quantities that
represent the tendency of fluid parcels to deform and rotate,
respectively.

One of the most prominent processes occurring at small scales is
the so-called `vortex stretching': following a common
argument,~\cite{tennekes} if a vortical fluid element is stretched
by the surrounding flow, the rotation rate should increase to
conserve angular momentum. However, L\"{u}thi et al.~\cite{luethi}
showed that this does not hold true point-wise and the dynamics
are significantly influenced by a viscous contribution. The
enstrophy balance equation,
\begin{equation}\label{eq1}
\frac{D}{Dt}\frac{\omega^2}{2}=\omega_i \omega_j s_{ij}+\nu
\omega_i \nabla^2 \omega_i,
\end{equation}
where the squared vorticity magnitude $\omega^2$ denotes the
enstrophy, $s_{ij}$ the rate of strain tensor and $\nu$ the
kinematic viscosity of the fluid, contains a production term \oos\
and a viscous term \odo. The two terms in the mean (hereinafter
mean values $< \cdot >$ are obtained by spatial and temporal
averaging) approximately balance each other, i.e.,
$\langle$\oos$\rangle \simeq -\langle$\odo$\rangle$, see
Ref.~\onlinecite{tennekes}. The presence of a viscous contribution
in Eq.~(\ref{eq1}) shows that the effect of molecular viscosity is
not limited to energy dissipation through deformation work,
expressed as $\varepsilon=2\nu s_{ij}s_{ij}$, but, among other
things, it controls also vorticity growth. The effects of vortex
stretching and viscous destruction are usually captured in the
well-known picture that in turbulence at small scales the
nonlinearities increase gradients, whereas the viscosity depletes
them, e.g. Refs.~\onlinecite{tennekes},\onlinecite{tsinober} and
references therein. However, as noted already by, e.g. Tennekes
and Lumley,\cite{tennekes} viscous effects are not restricted to
vorticity destruction only. For example, viscosity may tilt
vorticity, see, e.g.
Refs.~\onlinecite{tennekes},\onlinecite{tsinober},\onlinecite{kida},\onlinecite{takaoka2},\onlinecite{guala}
and is believed to be responsible for vortex reconnection, e.g.
Ref.s~\onlinecite{tsinober},\onlinecite{kida}
and~\onlinecite{takaoka2}. It is reminded that this `classical'
reconnection mechanism (due to viscosity) is fundamentally
different from reconnection events in quantum fluids, which take
place due to a quantum stress acting at the scale of the vortex
core without changes of total energy.~\cite{leadbeater,paoletti}
However, direct experimental evidence for the occurrence of
tilting and production of vorticity due to viscosity is still
missing in the literature, also because up to now it was difficult
to measure the associated small scale quantities experimentally.
Derivatives of the velocity became accessible through particle
tracking experiments since the developments in, e.g.
Ref.s~\onlinecite{luethi},\onlinecite{voth},\onlinecite{holzner}.
Holzner et al.~\cite{holzner} recently measured viscous production
of vorticity in proximity of turbulent/nonturbulent interfaces,
which raised the question about the role of positive \odo\ in
fully developed and homogeneous turbulence.\\In this letter we
present the first measurements of tilting, depletion and
considerable production of vorticity through viscosity in a
turbulent flow through particle tracking velocimetry
(Ref.s~\onlinecite{luethi},\onlinecite{voth},\onlinecite{holzner}).
The main goal is to unfold viscous effects on vorticity dynamics
at the small scales of turbulence, with an emphasis on genuine
(i.e. intrinsic to Navier Stokes turbulence as opposed to
kinematic) effects. The results discussed hereafter are based on
higher order derivatives and are challenging to obtain, both
experimentally and numerically, which is why we compare the
experimental results with those obtained through direct numerical
simulation.

We measured the flow velocities and its gradients in a laboratory
experiment of homogeneous, quasi isotropic and statistically
stationary turbulence by using particle tracking velocimetry, see
Ref.s~\onlinecite{luethi} and~\onlinecite{hoyer} for details.
Particle tracking velocimetry is based on high speed imaging of
the motion of small buoyant tracer particles seeded into the flow.
The experiment was carried out in a glass tank filled with water
and the flow was forced mechanically from two sides by two sets of
rotating disks as in Ref.~\onlinecite{hoyer}. The observation
volume of approximately 15 x 15 x 20 mm$^{3}$ was centered with
respect to the forced flow domain, mid-way between the disks. The
turbulent flow is characterized by an r.m.s velocity of about $10$
mm/s, a Taylor-based Reynolds number of Re$_{\lambda }=50$ and the
Kolmogorov length and time scales are estimated at $\eta=$0.5~mm
and $\tau_{\eta}=$0.25~s, respectively. The Laplacian of
vorticity, $\nabla^2\mbox{\boldmath{$\omega$}}$, is obtained
indirectly from the local balance equation of vorticity in the
form $\nabla \times\mathbf{a}=\nu
\nabla^2\mbox{\boldmath{$\omega$}}$ by evaluating the term $\nabla
\times \mathbf{a}$ from the Lagrangian tracking data. Through this
indirect method only one derivative in space is needed instead of
three, but particle positions have to be differentiated twice in
time in order to get Lagrangian acceleration. For the numerical
simulation we used an open source turbulence
database~\footnote{http://turbulence.pha.jhu.edu/} that was
developed at Johns Hopkins University, see
Ref.s~\onlinecite{perlman},\onlinecite{li} for details. The data
are from a direct numerical simulation of forced isotropic
turbulence on a $1024^3$ periodic grid, using a pseudo-spectral
parallel code. The Taylor Reynolds number is $Re_{\lambda}=434$.
After the simulation had reached a statistically stationary state,
$1024$ frames of data, which includes the $3$ components of the
velocity vector and pressure, were generated and stored into the
database. The time interval covered by the numerical data set is
thus only one large-eddy turnover time, whereas it is $O(10)$
turnover times for the experiment. For comparison to a random
velocity field, divergence-free Gaussian white noise was generated
as in Ref.~\onlinecite{bos}.


\begin{figure}
\begin{center}
\includegraphics[width=.5\textwidth,keepaspectratio=true]{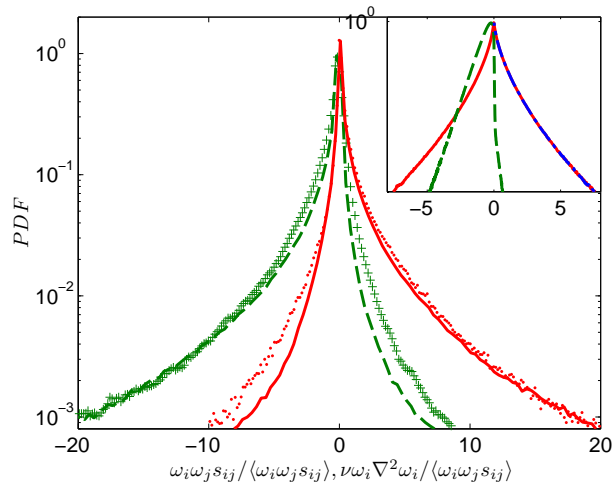} 
\end{center}
\caption{$PDFs$ of \oos\ (---, $\bullet$) and \odo\ ($-\ -$, $+$)
normalized with $\langle$\oos$\rangle$. Symbols are from PTV,
lines from DNS. The inset shows the analogous results from a
random Gaussian velocity field, \oos\ (---), $\omega_i
\nabla^2\omega_i$ ($-\ -$), the vertical reflection of the $PDF$
corresponding to negative events, \oos$<0$ ($-\ \cdot\ -$),
demonstrates the symmetry.\label{fig:1}}
\end{figure}
First, we statistically analyze effects of viscosity on the
vorticity magnitude. One of the most basic phenomena of three
dimensional turbulence is the predominant vortex stretching, which
is manifested in a positive net enstrophy production,
$\langle$\oos$\rangle>$0, e.g.,
Refs.~\onlinecite{tennekes},\onlinecite{tsinober} and references
therein. A strong positive skewness of the Probability Density
Function ($PDF$) of the term \oos\ is indeed visible in
Fig.~\ref{fig:1}, in agreement with earlier results, e.g.
Ref.~\onlinecite{tsinober}. For statistically stationary
turbulence the growth of enstrophy is balanced by viscous effects,
i.e., the two terms on the RHS of Eq.~(\ref{eq1}) balance in the
mean. Consistently, the term \odo\ shows an opposite distribution,
being strongly negatively skewed (Fig.~\ref{fig:1}). Although
viscosity mostly depletes enstrophy, we note that also events
where \odo$>0$ are statistically significant. In fact, about one
third of all events represent viscous production of enstrophy. The
experimental curves qualitatively agree with the numerical ones,
the $PDFs$ obtained from DNS are slightly more skewed. It is
important to note that, while the reasons for the positiveness of
the mean enstrophy production term are dynamical and due to
interaction between vorticity and strain, the destructive nature
of the viscous term $\langle$\odo$\rangle<0$ arises also for
kinematical reasons: one can decompose the viscous term as, e.g.
\begin{equation}
\mathbf{\omega_i}\nabla ^{2}\mathbf{\omega_i} =
-\nabla\cdot(\mathbf{\omega }\times (\nabla\times\mathbf{\omega}))
-(\nabla\times\mathbf{\omega })^{2},
\end{equation}\label{eq:2}
where the first term on the RHS is a divergence of a vector and
vanishes in the mean for homogeneity, whereas the second is a
(always negative) dissipation term~\footnote{\scriptsize It is
noteworthy that the above decomposition of $\nu \omega _{i}\nabla
^{2}\omega _{i}$\ - though useful - has a limitation since it is
not unique and there is an infinite number of possibilities to
represent it as a sum of a dissipation and a flux term (i.e. as a
divergence of some vector). There is no way to define dissipation
(i.e. to choose one among many purely negative expressions) of
enstrophy as it is not an inviscidly conserved quantity, unlike
the kinetic energy.\cite{tsinober}}. Indeed, while for a Gaussian
random field $\langle$\oos$\rangle$=0 and the $PDF$ of \oos\
becomes symmetric, the $PDF$ of the viscous term is strongly
negatively skewed, see the inset in Fig.~\ref{fig:1}. This means
that the destructive nature of the viscous term is also recovered
in a random field and does not represent a genuine property of
turbulent flow fields. However, from the same inset, we estimate
that for a random gaussian field, the events with
$\omega_{i}\nabla ^{2}\omega _{i}>0$ are statistically far less
significant (about 2\% of all events) compared to the same events
in a Navier Stokes field (about 30\%). We therefore conclude that
considerable viscous production of vorticity is a genuine
characteristic of Navier Stokes turbulence.
\begin{figure}
\begin{center}
\includegraphics[width=.65\textwidth,keepaspectratio=true]{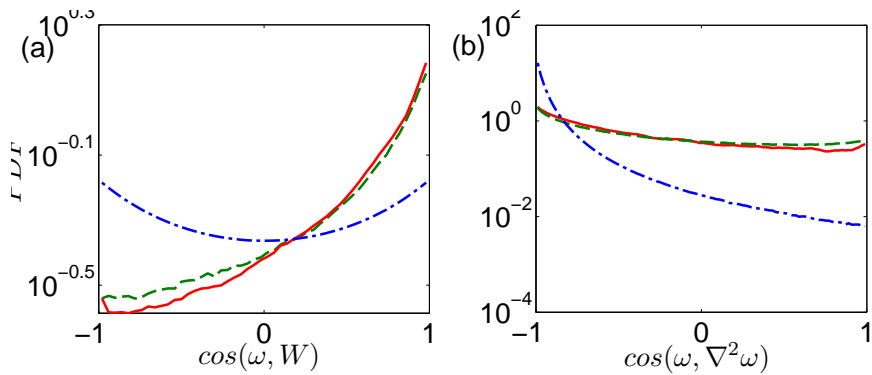} 
\end{center}
\caption{$PDFs$ of the cosine between vorticity and the vortex
stretching vector (a) and between vorticity and its Laplacian (b),
as obtained from DNS (---), PTV ($-\ -$) and random Gaussian field
($-\ \cdot\ -$).\label{fig:2}}
\end{figure}

The positiveness of the mean enstrophy production is associated
with the predominant alignment between vorticity and the vortex
stretching vector. The enstrophy production can be expressed as
the scalar product of vorticity and the vortex stretching vector,
\oos=\ome$\cdot\mathbf{W}$, where $W_i=\omega_js_{ij}$. In real
turbulent flows, the two vectors are strongly aligned. Thus, the
$PDF$ of the cosine between \ome\ and $\mathbf{W}$ is asymmetric
(Fig.\ref{fig:2}a), in conformity with the prevalence of vortex
stretching over vortex compression, whereas it is symmetric for a
random Gaussian field (Fig.~\ref{fig:2}a), see also
Ref.~\onlinecite{tsinober} and references therein. Analogously, we
show the alignment between \ome\ and \vis\ in Fig.~\ref{fig:2}b.
The figure shows high probabilities (much higher for the random
field) of pronounced anti-alignment between \ome\ and \vis,
consistent with the negative skewness of the $PDF$ of \odo, but we
also note that with some smaller probability the two vectors can
attain any orientation and, in particular, they can also be
strongly aligned. This reminds of the results in
Ref.~\onlinecite{holzner}, who measured cos$($\ome,\vis$)\simeq$ 1
in the proximity of the interface between turbulent and
irrotational flow regions. The fact that the two vectors are not
always strictly anti-aligned implies that the term $\nu$\vis\ does
not act exclusively in the direction of the vorticity vector
(mostly dampening and sometimes increasing the vorticity
magnitude), but also normally to it, thus contributing to altering
the orientation of vorticity. Since the negative skewness of the
$PDF$ is much stronger for the random velocity field than for the
turbulent one, we may infer that viscous tilting is characteristic
of fluid turbulence. The observation that the viscous term can
effectively influence the orientation of vorticity is important,
also because this will affect the relative orientation between
\ome\ and \lai\ and therefore indirectly influence the vortex
stretching (compression) mechanism.

\begin{figure}
\begin{center}
\includegraphics[width=.4\textwidth,keepaspectratio=true]{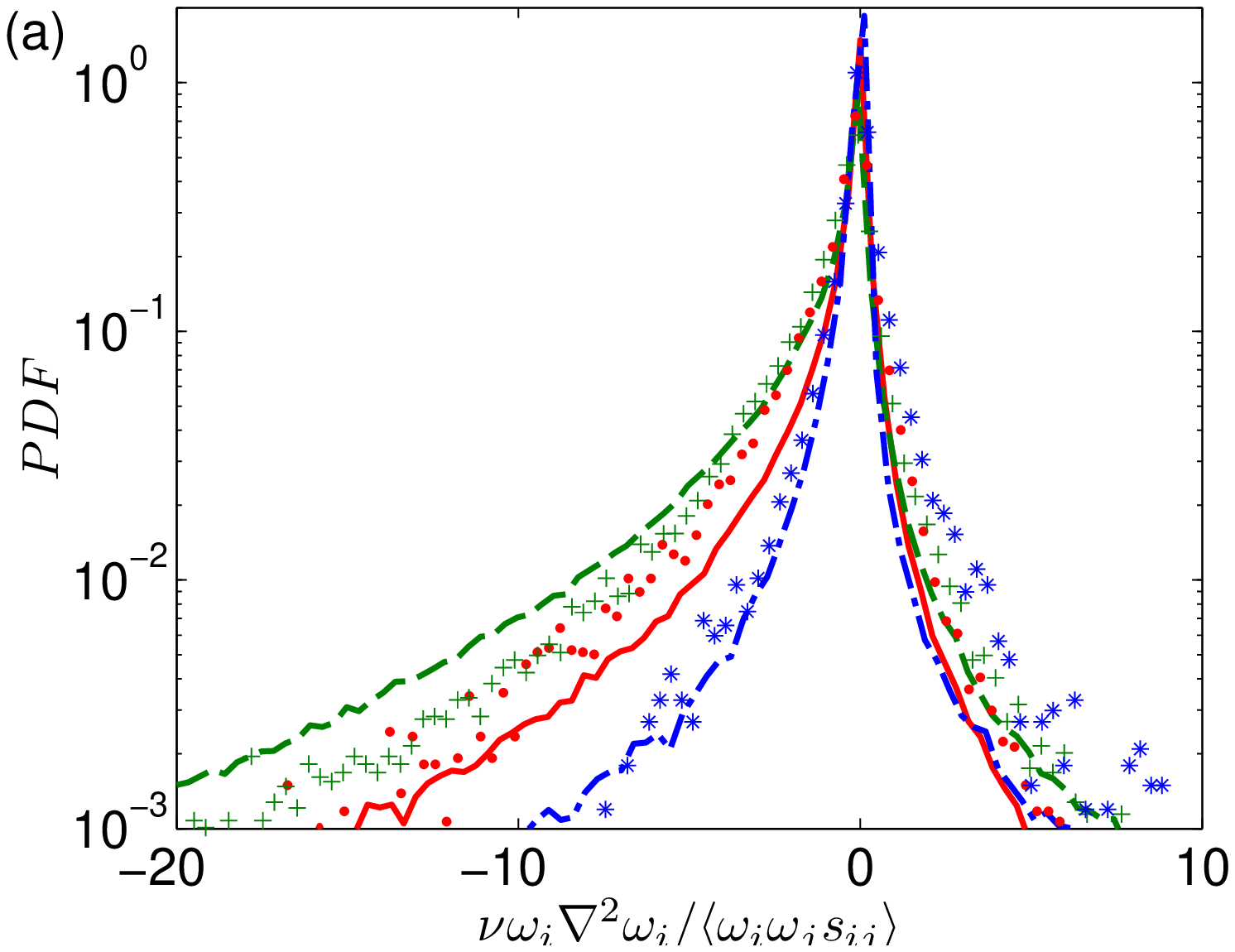} 
\includegraphics[width=.4\textwidth,keepaspectratio=true]{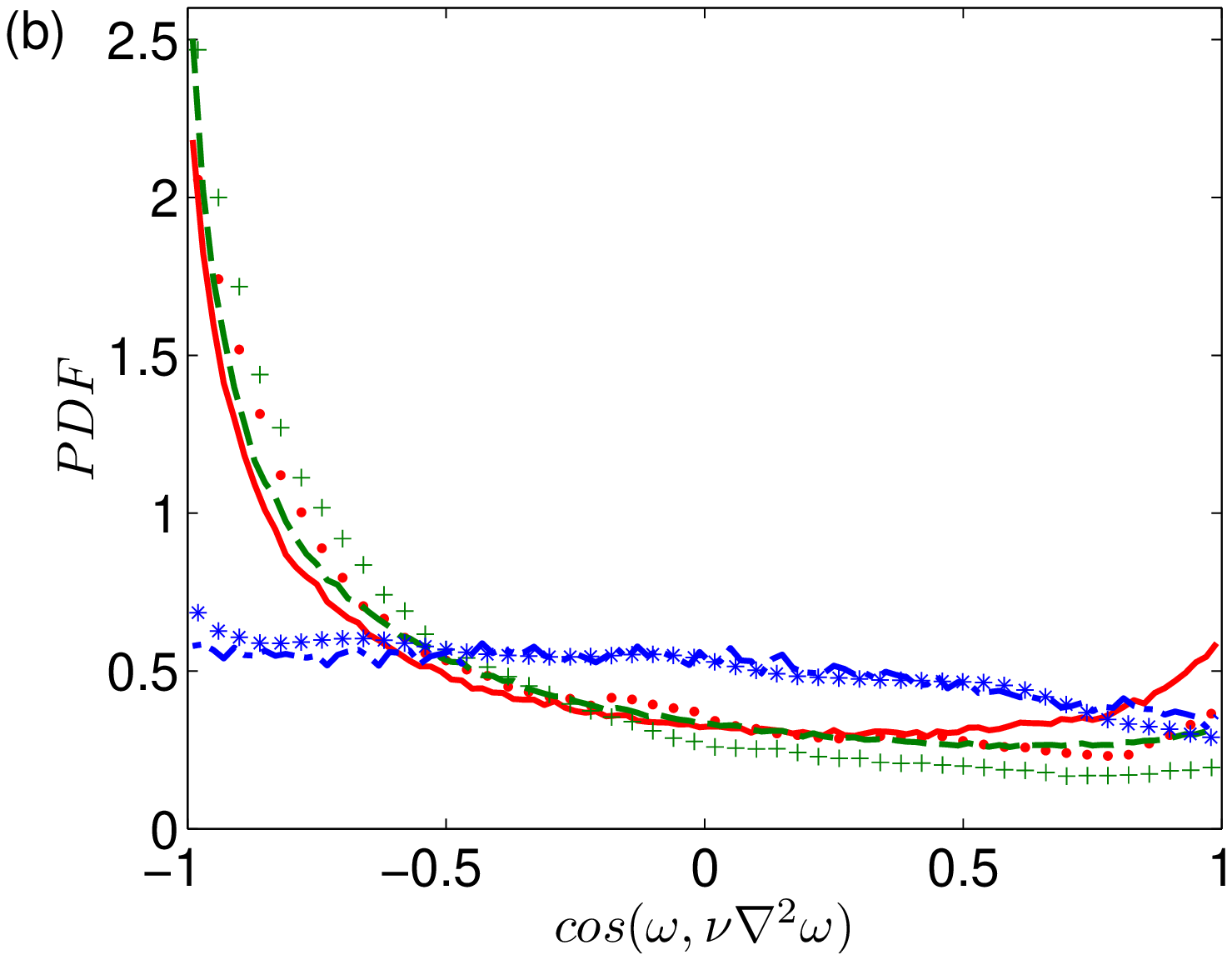}
\end{center}
\caption{$PDFs$ of \oos\ and \odo\ (a) and of the cosine between
vorticity and its Laplacian (b) for different (\ome-\lai)
alignments from DNS (lines) and PTV (symbols), \ome\ aligned with
\laa\ (---, $\bullet$), \lab\ ($-\ -$ , +) and \lac\ ($-\ \cdot\
-$, $\ast$).\label{fig:3}}
\end{figure}

The inviscid tilting of vorticity was measured by Guala et
al.\cite{guala} and found to be sensitive to the alignments
between vorticity and the strain eigenvectors. With the present
data it is possible to estimate for the first time both the
inviscid and the viscous contribution to the tilting of vorticity
and to quantify the influence of the relative (\ome-\lai)
alignments. We adopt the approach of Ref.~\onlinecite{guala} and
condition the data on situations of different alignment of
vorticity with the principal axis of the strain eigenframe. Note
that in a Gaussian field no differences are observed when
conditioning on such alignments and therefore the expected effects
in turbulent flow are explicitly dynamical.

Fig.~\ref{fig:3}a depicts the $PDFs$ of the two terms divided into
the three subsets depending on the local alignment between
$\mbox{\boldmath{$\omega$}}$\ and $\mbox{\boldmath{$\lambda_i$}}$.
The subsets are divided according to the condition
$\cos^2($\mbox{\boldmath{$\omega$}}$,$\mbox{\boldmath{$\lambda_i$}}$)
\leq 0.7$, corresponding to a cone of roughly $33^{\circ}$, as in
Ref.~\onlinecite{guala}. It is visible that, while for the case of
alignment with the intermediate eigenvector, \lab, the $PDF$
becomes more skewed, i.e. \odo\ contributes more to the reduction
of $\omega^2$, whereas in the case of alignment with \laa, the
skewness decreases and even more so when vorticity is aligned with
\lac. Again, the main qualitative trends are the same both for the
numerical and experimental results, with the curves obtained from
DNS showing a stronger skewness.

In Fig.~\ref{fig:3}b we analyze how this qualitatively different
behavior of the term \odo\ is reflected in the alignment between
\ome\ and \vis. The $PDF$ of the cosine between the two vectors is
strongly negatively skewed for the cases when \ome\ is aligned
with \laa\ and \lab. In the case of \ome\ aligned with \lac\ the
distribution changes dramatically becoming very flat in conformity
with the reduced skewness of the $PDF$ of \odo. Therefore, in this
case viscosity contributes less to the destruction of enstrophy,
but still plays a role, e.g. for the tilting of the vorticity
vector.

\begin{figure}
\begin{center}
\includegraphics[width=.65\textwidth,keepaspectratio=true]{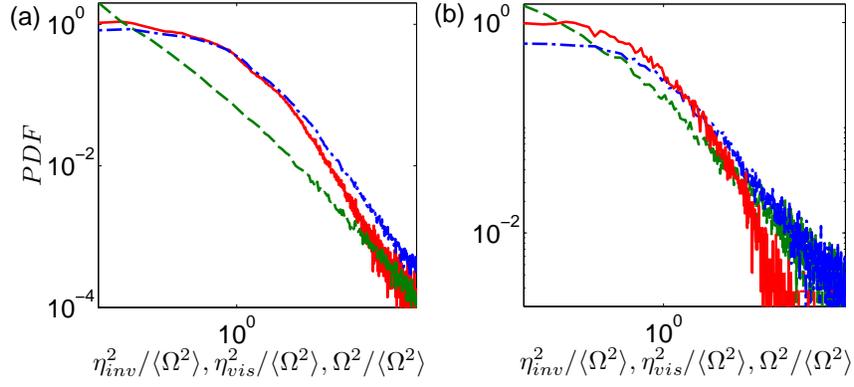}
\end{center}
\caption{PDFs of inviscid (---), viscous ($-\ -$) and total ($-\
\cdot\ -$) tilting from DNS (left) and PTV (right).\label{fig:5}}
\end{figure}

The inviscid and the viscous contribution to the total tilting
$\mathbf{\Omega}$ of vorticity can be written as follows,
\begin{equation}  \label{eq:til}
\Omega_{k}=\frac{D\widehat{\omega }_{k}}{Dt}=\eta _{\omega
_{k}}^{i}+\eta _{\omega _{k}}^{v},
\end{equation}
where $\eta _{\omega _{k}}^{i}=\frac{\omega _{j}s_{kj}}{\omega }-\frac{%
\omega _{l}\omega _{j}s_{lj}}{\omega ^{3}}\omega _{k}$ $\ $and
$\eta _{\omega _{k}}^{v}=\frac{\nu \nabla ^{2}\omega _{k}}{\omega
}-\frac{\nu \omega _{j}\nabla ^{2}\omega _{j}}{\omega ^{3}}\omega
_{k}$ represent the inviscid and viscous tilting respectively.
Fig.~\ref{fig:5} shows $PDF$s of the squared magnitudes of total,
inviscid and viscous tilting and it appears that viscous tilting
is typically smaller than the inviscid one, but at large
magnitudes both contributions to the total tilting are comparably
significant. The $PDFs$ of viscous and total tilting obtained from
PTV appear to be somewhat higher at the tails compared to the
numerical result, but the experimental scatter is considerable at
high magnitudes. In order to appreciate the dependence of the
tilting magnitudes on geometrical properties introduced before, it
is useful to write the following equations,
\begin{eqnarray}
(\eta_{\omega}^{i})^2&=&\sqrt{\mathbf{W}^2/\omega^2}sin^2(\mbox{\boldmath{$\omega$}},\mathbf{W})\\\label{eq:etai2_1}
&=&\Lambda_k^2cos^2(\mbox{\boldmath{$\omega$}},\mbox{\boldmath{$\lambda_k$}})-(\Lambda_kcos^2(\mbox{\boldmath{$\omega$}},\mbox{\boldmath{$\lambda_k$}}))^2\label{eq:etai2_2}
\end{eqnarray}
and
\begin{equation}\label{eq:etav2}
(\eta_{\omega}^{v})^2=\sqrt{(\nu\nabla^2\mbox{\boldmath{$\omega$}})^2/\omega^2}sin^2(\mbox{\boldmath{$\omega$}},\nabla^2\mbox{\boldmath{$\omega$}}).
\end{equation}
From Eq.~(\ref{eq:etai2_2}) one can see that the inviscid tilting
vanishes identically, when \ome\ is strictly aligned with \lai.
This alignment can then only be changed in two ways: through
viscous tilting and/or through a change of the orientation of the
strain eigenframe.

In summary, in this letter we have shown that viscosity in two
thirds of all events depletes enstrophy and that there is an
essential contribution of kinematic nature to this effect. Viscous
tilting and production of vorticity, which occur in one third of
all events, are instead characteristic features of turbulent
flows. Our results demonstrate that viscosity influences enstrophy
production by changing vorticity in magnitude and direction. The
observed effects are sensitive to the (\ome-\lai)\ alignments and
thus to the local vortex stretching (compression) regime. When
\ome\ is aligned with \lac\ the purely destructive contribution of
\odo\ is strongly suppressed. From the technical point we note
that the experimental and numerical results agree well with each
other on the qualitative level. Some quantitative discrepancies
might be attributed to the fact that experimental measurements are
affected by limited spatial resolution, noise and to the
difference in Reynolds numbers. Finally, we propose a plausible
postulate regarding the role of viscosity for the predominant
\ome-\lab\ alignment so typical for turbulent flows, e.g.
Refs.~\onlinecite{ashurst} and~\onlinecite{hamlington}. In these
situations the vectors \ome\ and \vis\ are predominantly
anti-aligned. Strong \ome-\lab\ alignment stalls inviscid tilting,
while the anti-alignment of \ome\ and \vis\ points to reduced
viscous tilting, i.e., both mechanisms could work towards
maintaining the alignment. In future work we hope to pursue these
questions that are intimately related to moderation of enstrophy
growth and to prevention of finite time singularities.~\cite{kerr}
This will also require to address the tilting mechanisms of the
strain eigenframe.

%


\end{document}